\documentclass[twoside,fleqn]{article}
\usepackage{epsf}
\usepackage{espcrc2}
\newcommand{\beq}{\begin{equation}}
\newcommand{\eeq}{\end{equation}}
\newcommand{\half}{\frac{1}{2}}

\newcommand{\tab}[1]{Table~\ref{#1}}
\newcommand{\epsfaxhaxhax}[3]{
        \centerline{
          \hspace{-20pt}
          \epsfxsize=130pt
          {\epsfbox{#1}}
          \hspace{-15pt}
          \epsfxsize=130pt
          {\epsfbox{#2}}
          \hspace{-10pt}
          \epsfxsize=130pt
          {\epsfbox{#3}}}
}
\newcommand{\epsfaxhax}[2]{
        \centerline{
          \hspace{10pt}
          \epsfxsize=115pt
          {\epsfbox{#1}}
          \hspace{-10pt}
          \epsfxsize=115pt
          {\epsfbox{#2}}}
}
\newcommand{\epsfaxhaxme}[2]{
        \centerline{
          \hspace{-20pt}
          \epsfxsize=115pt
          \epsfysize=130pt
          {\epsfbox{#1}}
          \hspace{-10pt}
          \epsfxsize=115pt
          \epsfysize=130pt
          {\epsfbox{#2}}}
}

\title{Heavy Meson Spectroscopy at $\beta=6.0$}
\author{Peter Boyle,
\address{Department of Physics and Astronomy, 
University of Edinburgh, Edinburgh EH9 3JZ, Scotland
}
UKQCD Collaboration
}
 
\begin{document}

\begin{abstract}
We present results of a quenched calculation of the
heavy-light and quarkonium spectrum using
the tadpole improved clover action. We resolve completely
the triplet $\chi$ P-states in quarkonium systems, and
obtain evidence for fine structure of the heavy-light P-states.
Approximate scaling of the hyperfine splittings is observed,
producing results that are significantly below experiment.
\end{abstract}

\maketitle

\section{Introduction}
\vspace{-.05in}
Charm physics continues to pose difficulties for lattice
QCD simulations. The systems are significantly relativistic
\cite{ChristineTsukuba,Trottier} causing problems for the NRQCD
approach, and have $am_Q \simeq O(1)$ causing significant discretisation
effects in the heavy Wilson quark approach. We present the 
results of a simulation using the tadpole improved clover action
and perform the analysis using the 
Fermilab \cite{kronfeld_mflgt} interpretation
of the heavy Wilson quark approach.
\vspace{-.05in}

\section{Simulation Details}
\vspace{-.05in}
The simulation was performed using 499 quenched gauge
configurations on a $16^3\times48$ lattice.
Five heavy quark masses and three light
quark masses, detailed in \tab{QuarkMasses},
were simulated using the
tadpole improved clover action, 
with $u_0 = 0.8778$ from the average
plaquette, and $C_{SW} = 1.47852$.
\begin{table}[hbt]
\vspace{-.3in}
\caption{$\beta=6.0$ simulated kappas}
\label{QuarkMasses}
\begin{tabular}{ccc}
\hline
$\kappa $  & $aM_{PS}$ & Fuzzing Radius\\
\hline
0.13856& 0.228(2)&6\\
0.13810& 0.293(1)&6\\
0.13700& 0.4135(10)&6\\
0.13000& 0.9283(7)&3\\
0.12600& 1.1618(6)&3\\
0.12200& 1.3755(6)&3\\
0.11800& 1.5751(6)&3\\
0.11400& 1.7644(6)&3\\
\hline
\end{tabular}
\vspace{-.2in}
\end{table}
Both local and fuzzed \cite{cmi_fuzz} operators were generated 
at source and at
sink. (Local) covariant derivative sources in each of the spatial directions
were used for the $\kappa=0.12600$ quark corresponding to $\kappa_{\rm charm}$,
allowing the operator for a $^3P_2$ state to be created
for combinations involving $\kappa=0.12600$ with each of the other
masses. The operators used are given in \tab{tab:operators}. 

\begin{table}[hbt]
\vspace{-.25in}
\caption{Meson operators}
\label{tab:operators}
\begin{tabular}{rrc}
\hline
State      & $J^{PC}$ & Operators \\
\hline
$^1S_0$    & $0^{-+}$ & $\bar{\psi} \gamma_5 \psi$           \\
$^3S_1$    & $1^{--}$ & $\bar{\psi} \gamma_i \psi$           \\
$^1P_1$    & $1^{+-}$ & $\bar{\psi} \sigma_{ij}\psi$  \\
$^3P_0$    & $0^{++}$ & $\bar{\psi} \psi$ \\
$^3P_1$    & $1^{++}$ & $\bar{\psi} \gamma_i \gamma_5 \psi$ \\
$^3P_2$    & $2^{++}$ & $\bar{\psi} \{\gamma_i  \Delta_i - \gamma_j \Delta_j \}\psi$ ~\emph{E} {\rm rep}\\ 
           &          & $\bar{\psi} \{\gamma_i  \Delta_j + \gamma_j \Delta_i \}\psi$ ~\emph{T} {\rm rep}\\
\hline
\end{tabular}
\vspace{-.3in}
\end{table}

An extensive analysis of correlated
double and single exponential fits 
to various smearing combinations was carried out, and the optimal
fitting approach selected for each channel in the light-light, heavy-light
and heavy-heavy sectors independently.

\vspace{-.05in}
\section{Fine Structure}
\begin{figure*}[htb]
\epsfaxhaxhax{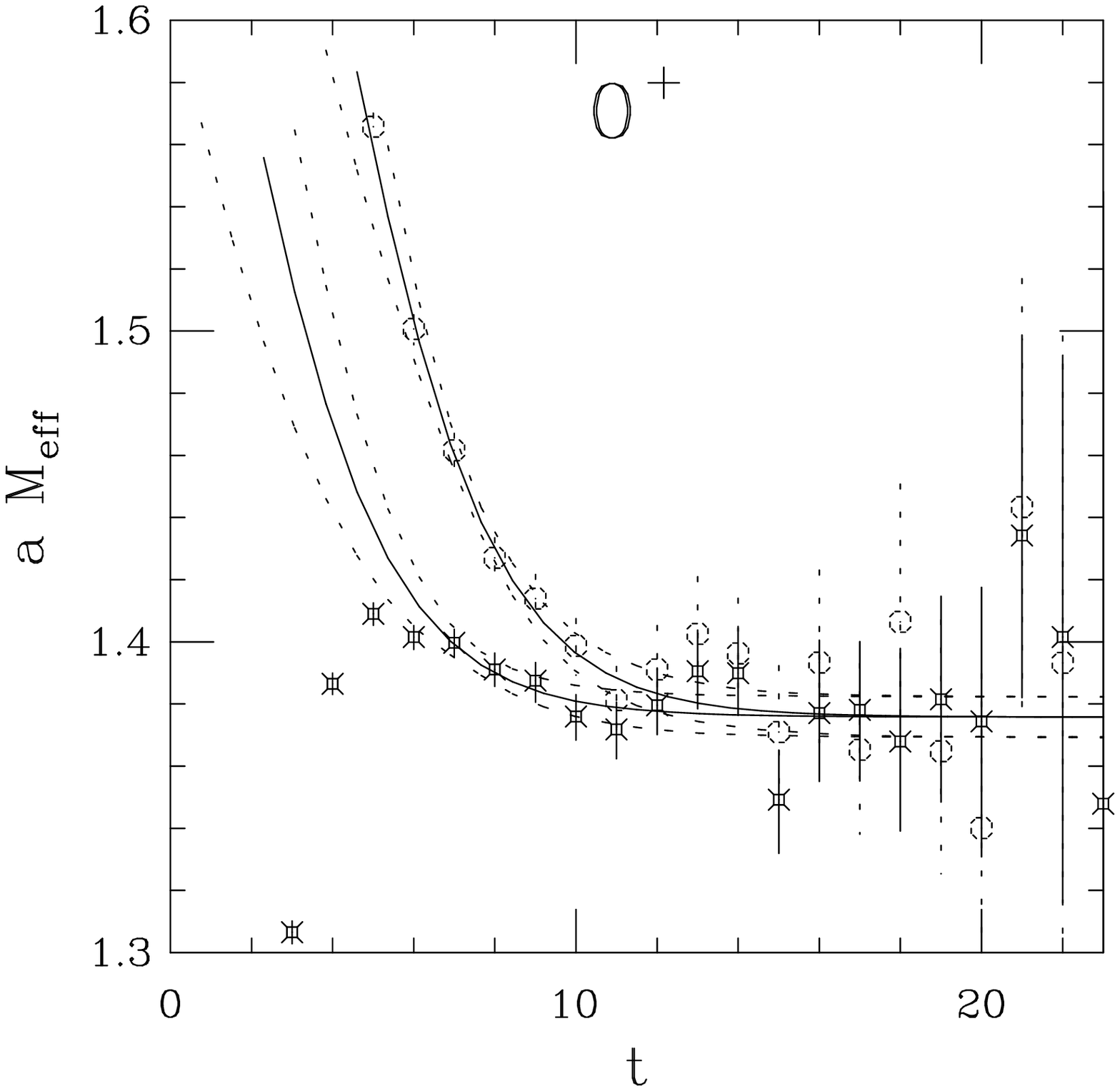}
{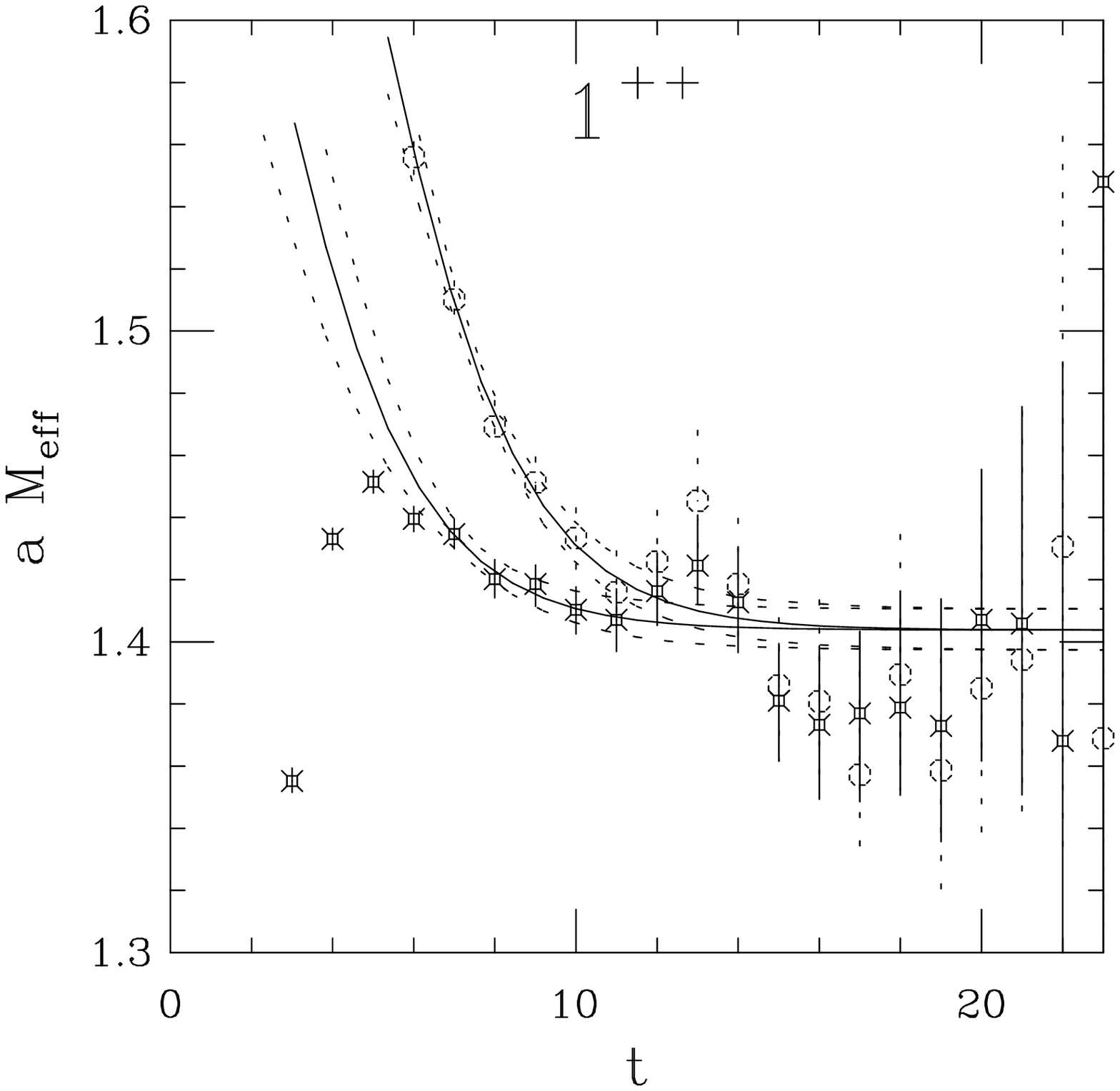}
{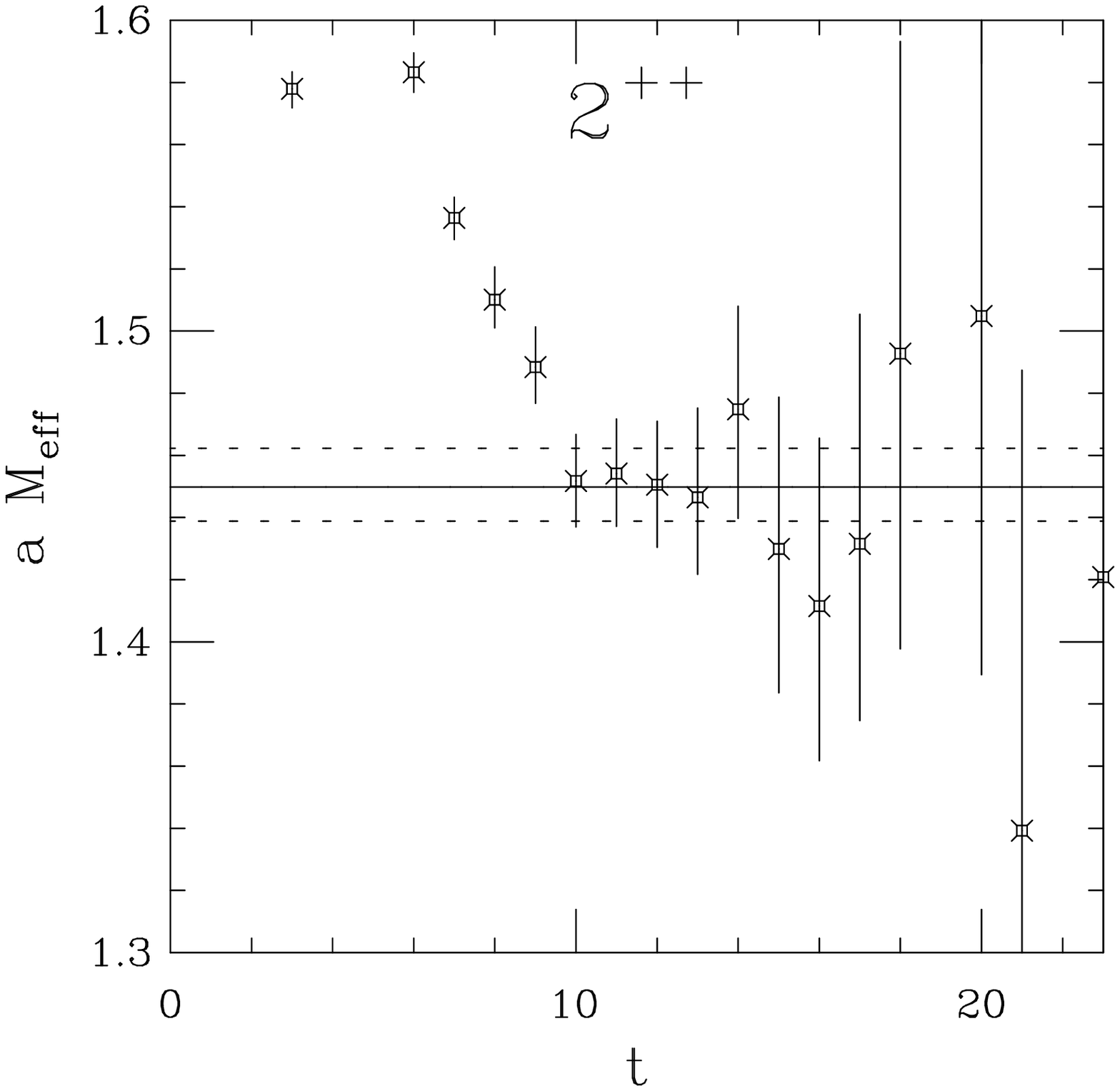}
\vspace{-0.2in}
\caption{
{\small
Quarkonium $\chi$ triplet for the $(\kappa=0.12600,\kappa=0.12600)$ combination.
Double exponential fits
were performed to the fuzzed-local and local-local correlators simultaneously
for the $0^{++}$ and $1^{++}$ states, while a single exponential fit
was performed to the derivative based operator for the $2^{++}$ state.
The three states are clearly resolved.
}}
\label{FigQQxc2}
\vspace{-0.25in}
\end{figure*}

\begin{figure}
\vspace{-0.3in}
\caption{
{\small Heavy-light fine structure for the
($\kappa_Q=0.12600$,$\kappa_q=0.13810$) combination; fit is to timeslices
7-13 of the fuzzed-fuzzed correlator.}}
\epsfaxhax{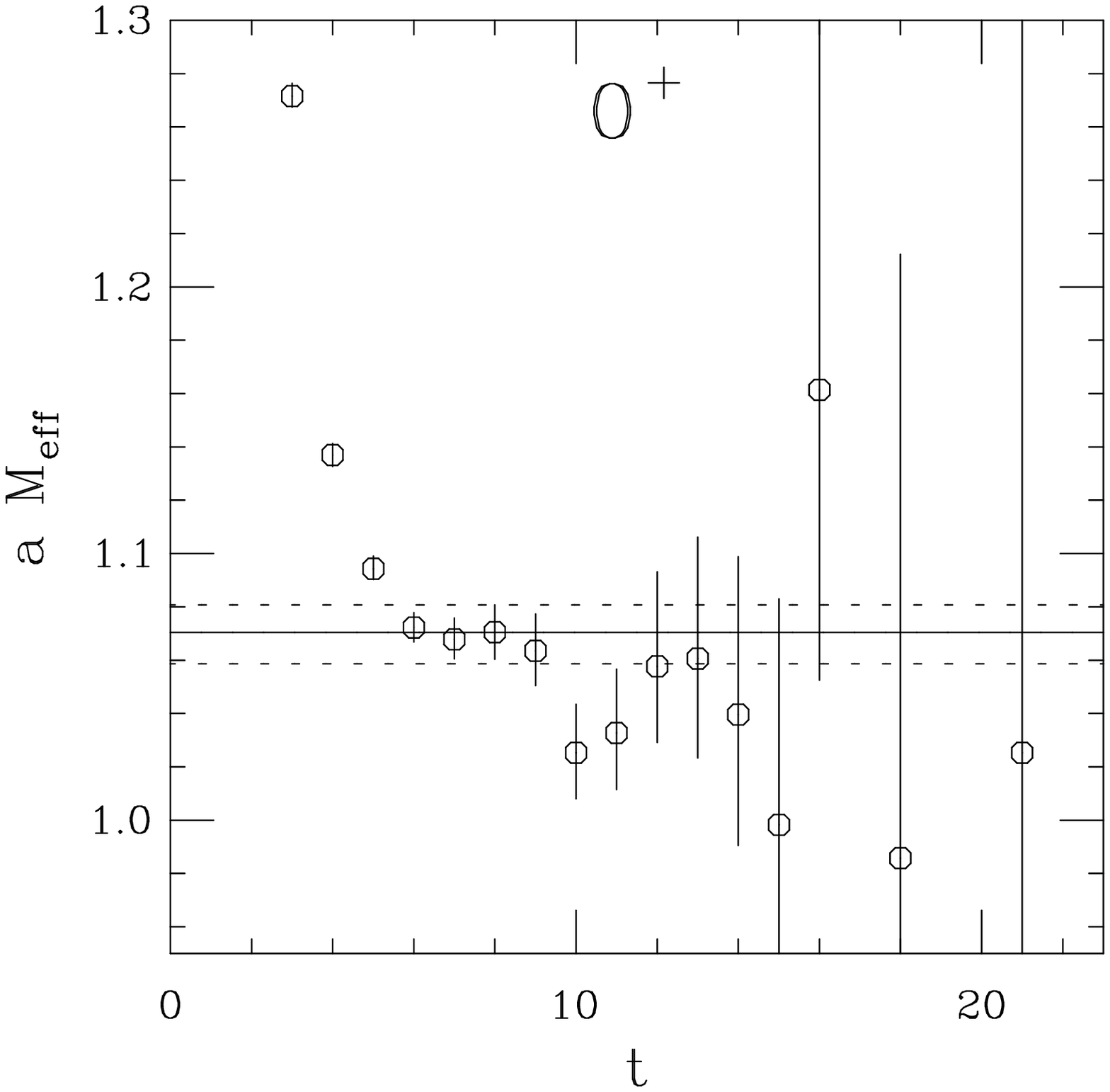}
{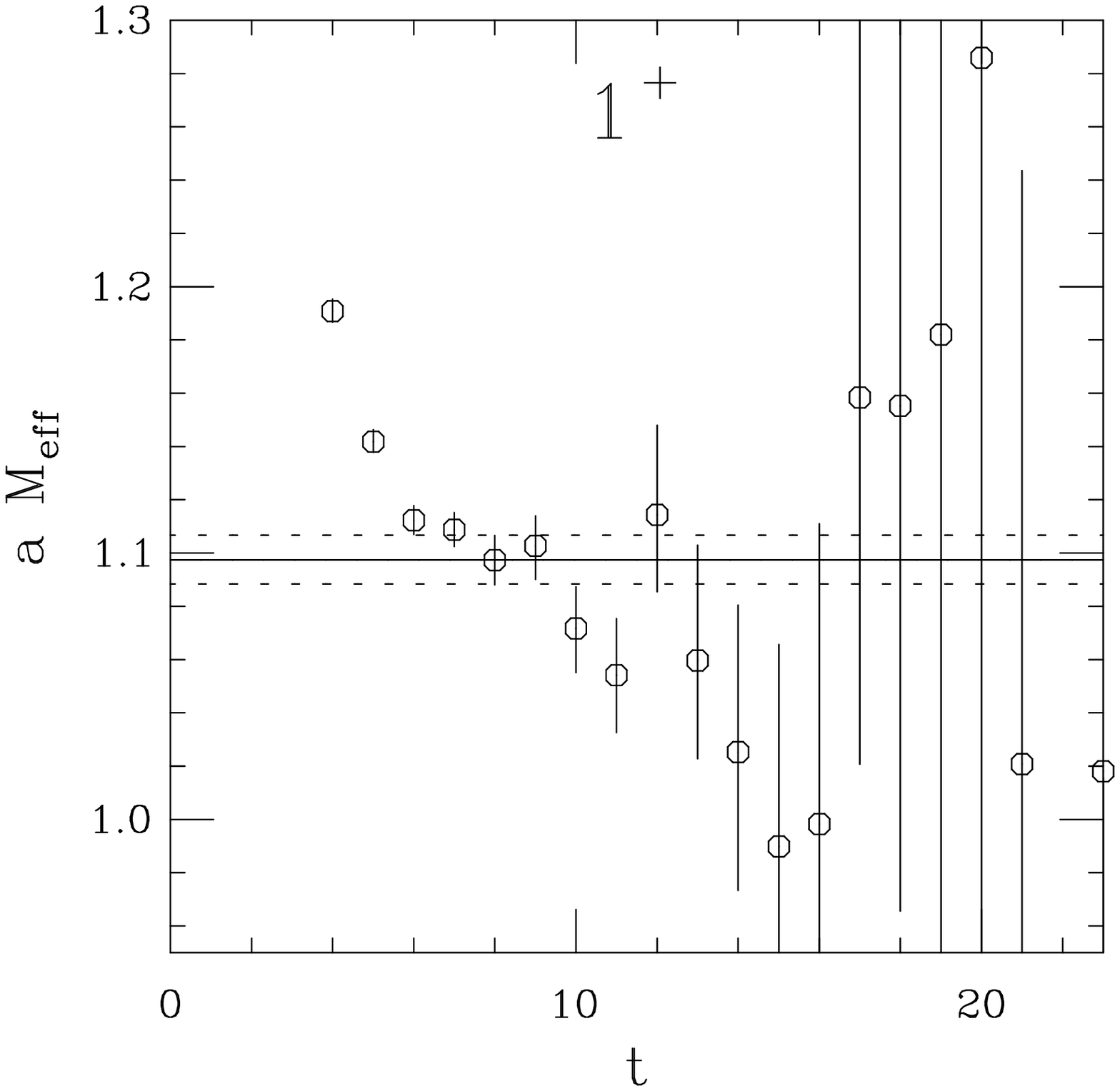}
\label{FigHL1+}  
\vspace{-0.2in}
\end{figure}

\vspace{-0.05in}
We obtain a signal for the fine structure of the 
$\chi$ triplet of P-states in quarkonium, as illustrated
in the effective mass plots in Figure \ref{FigQQxc2}. 
In the heavy-light case we obtain a signal for the splitting
between the $1^+$ and $0^+$ correlators, Figure \ref{FigHL1+},
fitted using a single exponential model
for the fuzzed-fuzzed
combination. The consistency of
both double and triple exponential
multi-correlator fits was checked.
Cross correlations between the $1^{++}$ and $1^{+-}$ operators
showed that there was significant mixing, as expected since
the $j_{\rm light}$ basis is the physical basis in the $m_Q\to\infty$ limit.
However we could not resolve the two physical $1^+$ states, and treat
the ground state for each of the $1^+$ correlators 
as the lower lying $1^+$ state, assumed to have $j_{\rm light}=\half$.

\vspace{-.05in}
\section{Spectrum at $\beta=6.0$}
\vspace{-.05in}
We use the dispersive mass $m_2$ of pseudoscalar mesons 
defined by
$
E(p^2) = m_1 + \frac{p^2}{2m_2} + C p^4,
$
as the definition of the heavy quark
mass in extrapolations of the spectrum.
We found that linear extrapolations in the inverse heavy quark mass
gave a very acceptable $\chi^2/{\rm dof}$.
For the heavy-light systems, linear chiral extrapolations
were performed to $\kappa_{\rm crit}$ and to $\kappa_{\rm strange}$,
followed by linear extrapolations in the inverse heavy-strange pseudoscalar
mass to the $D_s$ and $B_s$ masses.

For the $^3P_2$ state, where
only combinations of propagators involving the $\kappa=0.12600$ 
quark (with another) could be formed from the available data, 
the non-degenerate
combinations were used to extrapolate to the physical meson masses,
introducing a small correction in extrapolations to the $J/\psi$ system.
However, the extrapolations of the $^3P_2$ splittings
to $\Upsilon$ are not well under control.
The results obtained using 
the string tension, $m_\rho$ and the quarkonium $S-P$ splitting
to set the scale, 
and the kaon mass to fix 
$\kappa_{\rm strange}$, are tabulated in \tab{Tab60HLSpectrum}
\footnote{There is a systematic uncertainty of order 50 MeV \cite{Rosner}
in the values for the heavy-light S-P splitting since the experimental
values are for the $j_{\rm light} = \frac{3}{2}$ doublet, while
we calculate $j_{\rm light}=\frac{1}{2}$ states. For the 
heavy-light P-states we have adopted the nomenclature
used for the $j_{\rm light}=\frac{1}{2}$ doublet in the Kaon system by the
PDG.
}
and \tab{Tab60Quarkonium}.

\begin{table}[hbt]
\caption{$\beta=6.0$ heavy-light mass splittings (MeV)}
\label{Tab60HLSpectrum}
\begin{tabular}{ccccc}
\hline
Scale & $\sqrt{K}$ & $M_\rho$ & S-P &Expt \\
\hline
$D^* - D $              & 110(7)        & 106(8)        & 129(10) &142\\
$D^*_s - D_s$           & 99(5)         & 95(6)         & 115(9) &144\\
$\bar{D}_s - \bar{D}$     & 98(5)         & 96(5)         & 107(6) &105\\
$D_1 - \bar{D}$           & 540(30)       & 530(30)       & 600(30)&459\\
$D_{s1} - \bar{D_s}$      & 494(18)       & 480(20)       & 545(20)&460\\
$D_{1} - D_{0}^*$         & 45(20)        & 44(20)        & 47(25) &-\\
$D_{s1} - D_{s0}^*$       & 57(12)        & 56(11)        & 64(13) &-\\
\hline
$B^* - B $              & 41(9)         & 39(10)        & 59(11) &46\\
$B^*_s - B_s$           & 40(6)         & 38(7)         & 57(8)  &47\\
$\bar{B}_s - \bar{B}$     & 90(6)         & 88(6)         & 110(7) &91\\
$B_1 - \bar{B}$           & 490(30)       & 480(30)       & 590(40)&419\\
$B_{s1} - \bar{B_s}$      & 440(20)       & 430(20)       & 540(30)&446\\
$B_{1} - B_{0}^*$         & 50(20)        & 50(20)        & 60(20)&-\\
$B_{s1} - B_{s0}^*$       & 43(11)        & 42(10)        & 54(12)&-\\
\hline
\vspace{-.3in}
\end{tabular}
\end{table}

\begin{table}[hbt]
\vspace{-.1in}
\caption{$\beta=6.0$ heavy-heavy mass splittings (MeV)}
\label{Tab60Quarkonium}
\begin{tabular}{ccccc}
\hline
Scale & $\sqrt{K}$ & $M_\rho$ & S-P &Expt \\
\hline
$J/\psi - \eta_c$       &68(2)&65(2)&79(4)&117\\
$^1P_1 - \bar{S}$       &418(13)&408(16)&-&458\\
$\chi_{c2} - \chi_{c1}$ & 81(28)&78(27)&93(33)&46\\
$\chi_{c1} - \chi_{c0}$ & 51(7) &49(7)&59(11)&95\\
$\chi_{c2} - \chi_{c0}$ &133(28)&128(28)&153(35)&141\\
\hline
$\Upsilon - \eta_b$     &22(1)&21(1)&31(3)&40\\
$^1P_1 - \bar{S}$       &366(17)&358(20)&-&460\\
$\chi_{b2} - \chi_{b1}$ &43(30)&42(30)&56(32)&21\\
$\chi_{b1} - \chi_{b0}$ &26(9)&25(9)&33(10)&32\\
$\chi_{b2} - \chi_{b0}$ &65(30)&63(27)&85(32)&53\\
\hline
\end{tabular}
\vspace{-0.2in}
\end{table}

\vspace{-.05in}
\section{Scaling Behaviour}
\begin{figure}
\vspace{-0.3in}
\caption{Scaling Behaviour of hyperfine splittings.}
\epsfaxhaxme{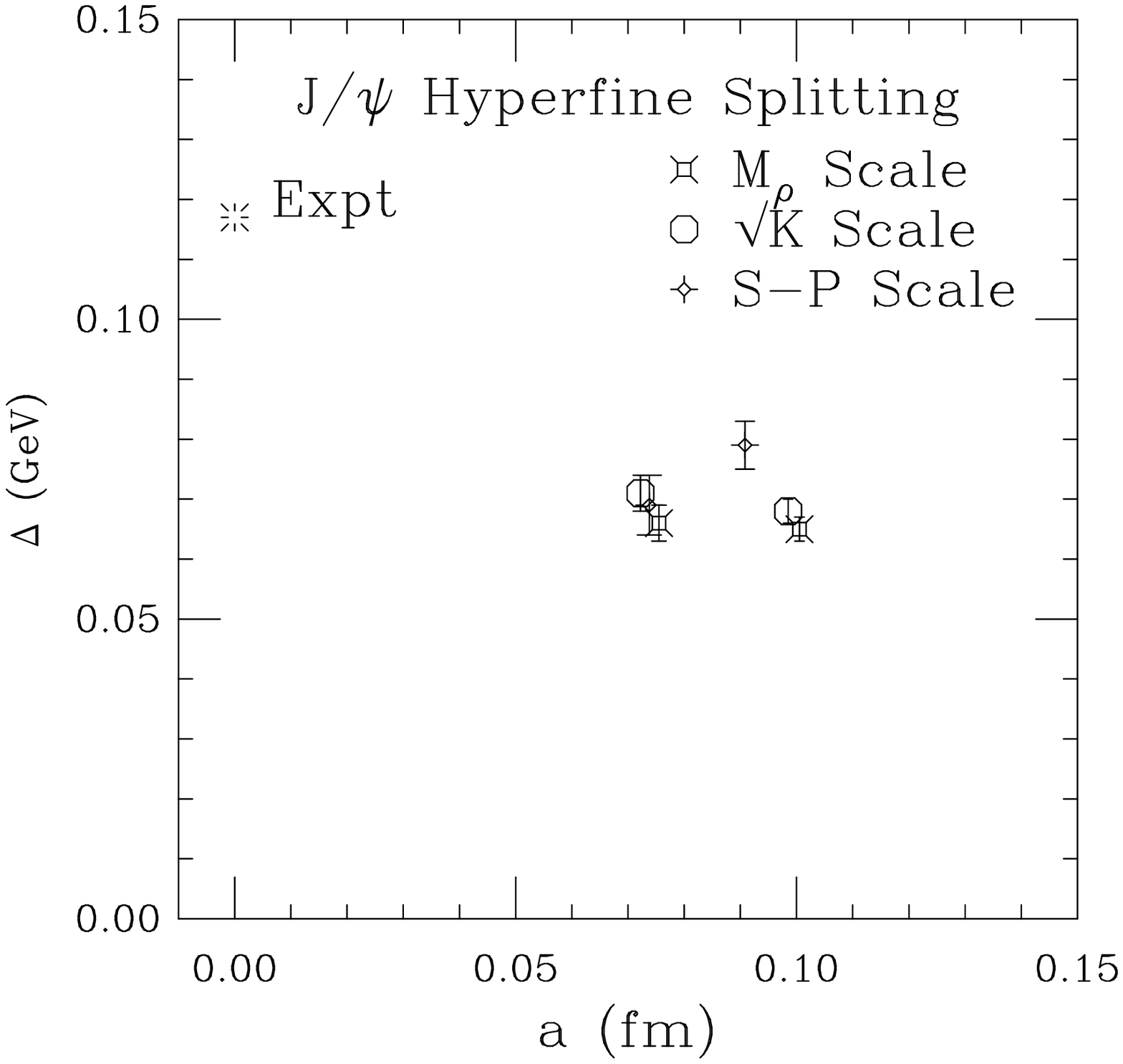}
{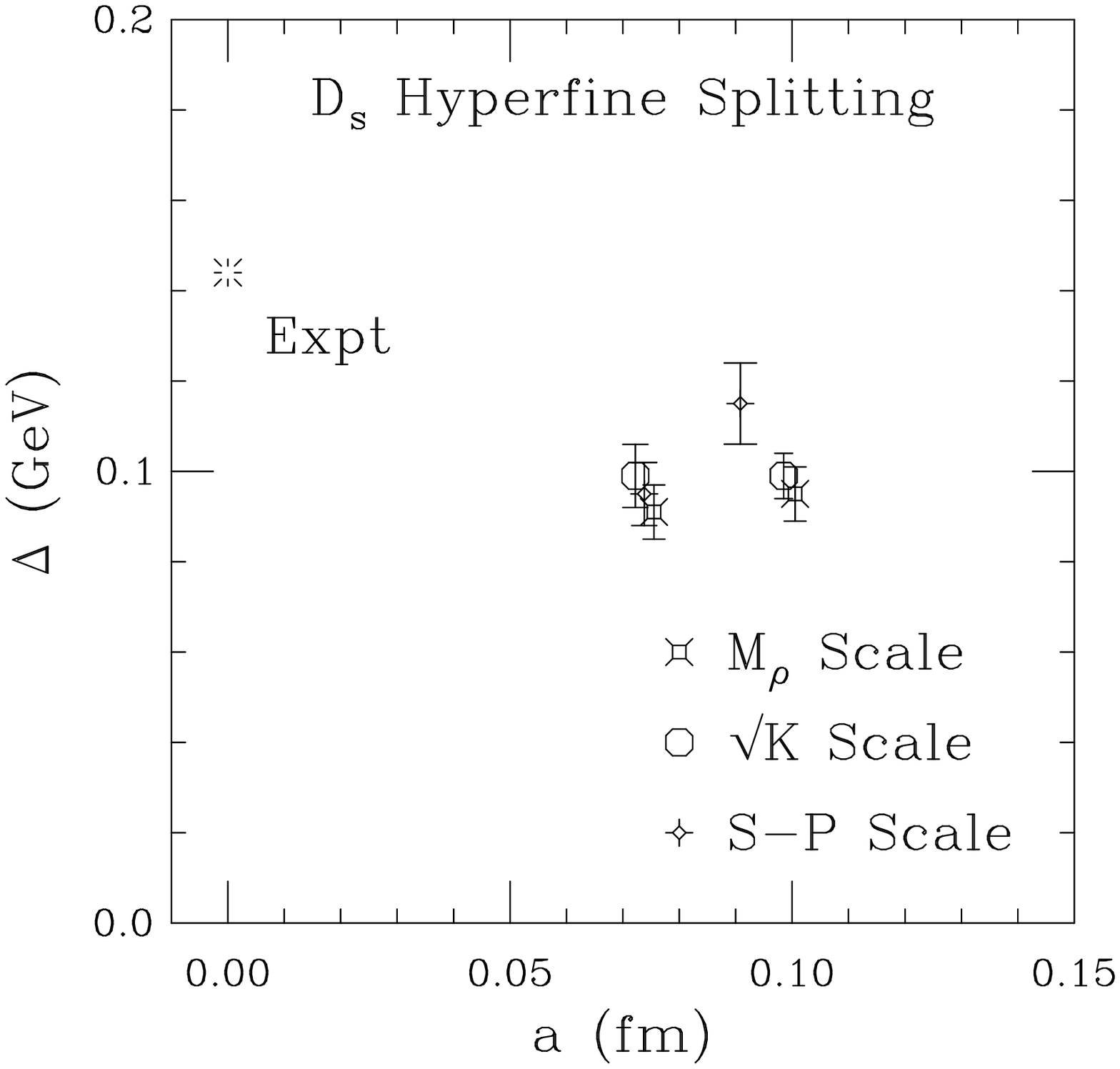}
\label{Scaling1}  
\vspace{-0.3in}
\end{figure}

\vspace{-.05in}
Comparison to a calculation using 220 configurations at
$\beta=6.2$ on a $24^3\times48$ lattice \cite{me_hee_hee}
with the same action, allows
some estimate of the scaling behaviour to be made. 
We plot the lattice spacing dependence of the charmonium and
$D_s$ hyperfine splittings in Figure \ref{Scaling1}.

Near-scaling behaviour is seen with both the string tension and
with $m_\rho$ used to set the scale. Scaling is not seen 
 with the quarkonium S-P
splitting. However, the lack of scaling is only a $1\sigma$ effect.
NRQCD calculations \cite{ChristineTsukuba} 
have found the $\Upsilon$ S-P splitting scaling well with $m_\rho$,
at about 30\% below its experimental value. This suggests that
ultimately scaling with respect to the S-P splitting will be seen 
at values of the hyperfine splittings above those with $m_\rho$
and the string tension. Likewise we find the $D$ and $\Upsilon$ hyperfine
splittings to show approximate scaling below experiment with 
$m_\rho$ and the string tension.
After extrapolating the heavy-light results to $B$ and $B_s$,
the errors are such that our values are consistent with experiment.

\vspace{-.05in}
\section{Conclusions}
\vspace{-.05in}
We resolve completely the triplet of $\chi$ states in the 
$J/\psi$ system with a relativistic action, and obtain evidence
for fine structure in heavy-light systems.
We find that both the quarkonium and D hyperfine splittings scale well with 
both the string tension and with $m_\rho$, lying significantly below their
experimental values. We find that our results using the quarkonium
S-P splitting do not scale well.
This may be due to the use of only
local propagators in the quarkonium calculation at $\beta=6.2$,
(which plateau at large $t$, making the plateau identification and statistical noise
toublesome), 
or to discretisation
effects. Further calculations 
(better smearing at $\beta=6.2$ and possibly at higher $\beta$)
are required to resolve which is the case.

\section*{Acknowledgements}
We acknowledge the support of EPSRC grant  GR/K41663 and PPARC
grant GR/K55745.
I was funded by the Carnegie Trust for the Universities of Scotland.
All calculations were performed using UKQCD time on the Cray T3D in
Edinburgh University.


\begin{thebibliography}{1}
 
\bibitem{ChristineTsukuba}
C.~Davies.
\newblock {\em hep-lat}, {\bf 9705037}, (1997).
 
\bibitem{Trottier}
H.~Trottier.
\newblock {\em Phys. Rev.}, {\bf D55}, 6844, (1997).
 
\bibitem{kronfeld_mflgt}
A~El-Khadra et. al.
\newblock {\em Phys. Rev. D}, {\bf 55}, 3933, (1997).
 
\bibitem{cmi_fuzz}
P.~Lacock et~al.
\newblock {\em Phys. Rev. D}, {\bf 51}, 6403, (1995).
 
\bibitem{Rosner}
J.~Rosner.
\newblock {\em Comm. Nucl. Part. Phys.}, {\bf 16}, 109, (1986).
  
\bibitem{me_hee_hee}
P.~Boyle.
\newblock {\em Nucl. Phys.}, {\bf B53 (Proc Suppl)}, 398--400, (1997).
 
\end{thebibliography}

\end{document}